# A hands-on Assessment of Transport Protocols with Lower than Best Effort Priority

Giovanna Carofiglio*, Luca Muscariello†, Dario Rossi‡ and Claudio Testa‡
* Bell Labs, Alcatel-Lucent, France, giovanna.carofiglio@alcatel-lucent.com
† Orange Labs, France, luca.muscariello@orange-ftgroup.com
‡ Telecom ParisTech, France, firstname.lastname@enst.fr

*Abstract*—Last year, the official BitTorrent client switched to LEDBAT, a new congestion control algorithm targeting a lower-than Best Effort transport service. In this paper, we study this new protocol through packet-level simulations, with a special focus on a performance comparison with other lower-than Best Effort protocols such as TCP-LP and TCP-NICE: our aim is indeed to quantify and relatively weight the level of Low-priority provided by such protocols.

Our results show that LEDBAT transport generally achieves the lowest possible level of priority, with the default configurations of TCP-NICE and TCP-LP representing increasing levels of aggressiveness. In addition, we perform a careful sensitivity analysis of LEDBAT performance, by tuning its main parameters in both inter-protocol (against TCP) and intra-protocol (against LEDBAT itself) scenarios. In the inter-protocol case, although in case of misconfiguration LEDBAT competes as aggressively as TCP, however we show that it is not possible to achieve an arbitrary level of low-priority by merely tuning its parameters. In the intra-protocol case, we show that coexistence of legacy flows with slightly dissimilar settings, or experiencing different network conditions, can result in significant unfairness.

## I. Introduction

BitTorrent, undoubtedly one of the most successful P2P filesharing applications nowadays, has recently adopted a new closed-loop congestion control algorithm, namely Low Extra Delay Background Transport (LEDBAT) [1], which is implemented at the application-layer and exploits UDP at the transport layer. The aim of the new protocol is "to not disrupt Internet connections, while still utilizing the unused bandwidth fully" [2] or in other words to delivery data with a lower priority in respect to general Best Effort, and thus TCP, traffic.

Lower than Best-Effort (LBE) priority is achieved in LED-BAT by reacting earlier than TCP to congestion notification, and reducing its transmission rate so to avoid harming TCP traffic: while TCP Reno infers congestion from packet losses, LEDBAT infers congestion from increasing buffering delay, hence prior than losses occur.

Thus, a first important observation about LEDBAT is that it constitutes a relief for operators, as they no longer need throttling the now gentle P2P traffic [3]. An additional relevant motivation for LEDBAT is that it relieves self-induced congestion when the bottleneck is placed at the user access link (e.g., DSL or cable). Self-induced congestion arises when users run in parallel several applications having different QoS constraints (e.g., Web browsing, gaming, VoIP, filesharing, backup): in this case, as the bottleneck is at the access, users are themselves generating competing traffic, but at the same time they would likely not want large background transfer to interfere with foreground interactive applications. In this context, LBE is a promising end-to-end technique that do not require coordination among applications, nor complex queuing policies or IP table rules to be setup by the end-users on their own PC [4].

However, many other services beside P2P filesharing may successfully exploit a LBE transport protocol as, e.g., the class of high volume data exchange, data mirroring and pre-fetching, network backups, etc. As such, LEDBAT is not the sole example of LBE transport that has been proposed in the literature: other notable protocols are for instance TCP-LP [5], TCP-NICE [6], 4CP [7] and Microsoft BITS [8]. Despite the relevance of the above scenarios, to the best of our knowledge no comparison attempt has been made yet between the different low-priority protocols: this situation is unlike the high-speed data transfer scenario, where several work [9], [10], [11] that compares different flavors of TCP exists, showing their relative merits and disadvantages.

In this work, we carry out a comparison of LBE protocols by means of `ns2` simulations, aiming at quantifying and ranking the relative level of LBE priority. We perform a systematic evaluation of the fairness and efficiency of three LBE protocols: namely, the new BitTorrent protocol LEDBAT [1], LP [5] and NICE [6]. Notice that only LP implementation is available as open source, so we implement both NICE and LEDBAT in the `ns2` simulator, and make our code available at [12]. As a scenario for the comparison, we consider the typical situation with many concurrent P2P flows sharing an access bottleneck link with other higher-priority traffic. Our results show that (i) LEDBAT transport achieves the lowest possible level of priority among the considered protocols, while NICE and LP represent increasing levels of aggressiveness. Moreover, we find that (ii) the level of low priority in LEDBAT cannot be easily set by tweaking the protocol parameters, and that (iii) in case of legacy LEDBAT implementations sharing the same bottleneck, even small differences in parameter settings (e.g., target delay) or network conditions (e.g., RTT delay) can result in significant unfairness.

## II. Lower than Best Effort transport Protocols

This section provides an overview of the relevant related work. On one hand, we have literature on BitTorrent, that until very recently [13], [14] has however focused on other aspects that LEDBAT, such as modeling BitTorrent performance [15], studying incentive mechanism [16] and locality-aware peer selection strategies [17], or analyzing torrent popularity [18].

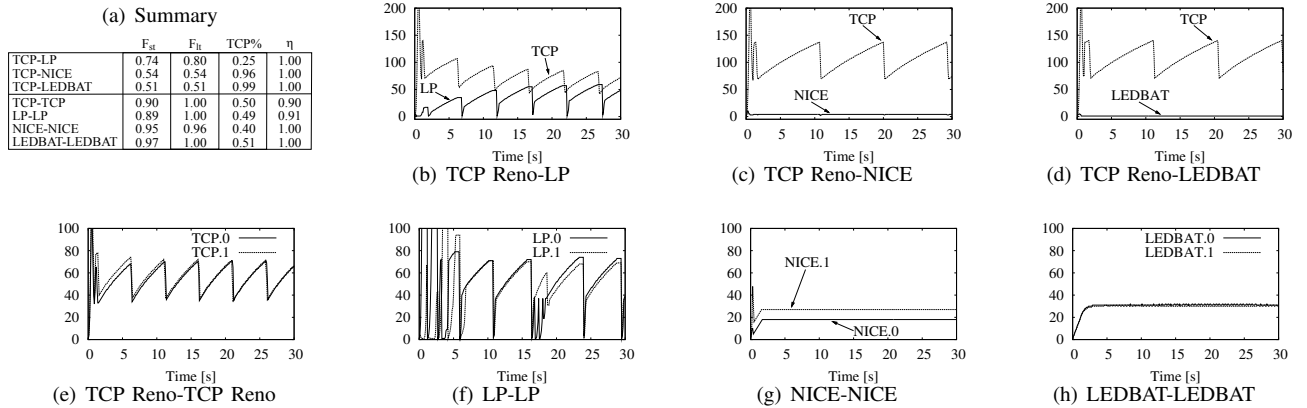

Fig. 1. Low priority at a glance: Inter (top) and Intra (bottom) protocol interaction on a simple bottleneck

On the other hand, we have studies that focus on congestion control: as the literature dates back to the late 80s [19] it is thus very wide. As such, we will especially focus on the LBE protocols that are considered in this work: namely, LEDBAT, LP and NICE, providing a detailed introduction that will be instrumental to the comparison. With this regard, we just point out that although work exists that, in similar spirit to ours compares high-speed TCP versions [9], [10], [11], to the best of our knowledge such a comparison effort has not been done for lower-than best effort protocols.

To achieve their goals, all LBE protocols need to detect congestion earlier than standard loss-based TCP. As the latter detects congestion by inferring that a packet loss occurred (e.g., by expiration of a timer, or by reception of duplicated acknowledgement), LBE protocols need to rely on a finer measure of congestion: typically, they equate increasing delay with incipient congestion. In other words, LBE protocols perform some delay measurement $D(t)$, and infer from the increase of $D(t)$ that congestion is building up, accounting the delay growth to some amount of queuing in the bottleneck link along the path. As we will see, the specific delay measurement $D(t)$ and the rule to decide that variations in $D(t)$ are actually due to congestion, vary from protocol to protocol. Then, once congestion has been (early) detected, this triggers a congestion-relief reaction, which again differs across protocols. It is however possible that congestion is not detected in a timely fashion, causing a packet loss of the LBE protocol: in this case, the reaction of all protocols falls back to standard TCP timeout mechanism, i.e. a drastic reduction of the congestion window $cwnd$.

In the rest of this section, we provide further details concerning each of the considered LBE protocols. To facilitate their comparison, we also report simple simulation results in Fig. 1, so to better visually highlight the relevant characteristics of each protocol. Top row of Fig. 1 reports the heterogeneous case where two flows employing different congestion control protocols are compared; bottom plot Fig. 1 shows time evolution of two flows employing the same LBE protocol assuming similar network conditions. More precisely, for each LBE$\in \{LP, NICE, LEDBAT\}$ protocol, Fig. 1 depicts the temporal evolution of the $cwnd$ of different flows in two scenarios of a simple bottleneck topology. In the inter-protocol scenario (top row, labeled as TCP-LBE), low-priority protocols compete against a standard TCP flow, while in the intra-protocol case (bottom row, labeled as LBE-LBE) two LBE flows compete against each other. In the figure, bottleneck capacity is set to C=10 Mbps, round-trip delay to RTT=50 ms and the buffer size is B=100 MTU-size packets.

### A. TCP-LP

TCP-LP (or LP *tout court*) measures one-way packet delays and employs a simple delay threshold-based method for early inference of congestion. More specifically, LP estimates the minimum $D_{min}$ and maximum one-way delay $D_{max}$, filtering the instantaneous measure of the delay $D(t)$ by means of an exponentially weighted moving average $\tilde{D}(t)$ with smoothing parameter $\alpha$, updated packet-by-packet. The smoothed average $\tilde{D}(t)$ and the condition for early congestion detection are:

$$\tilde{D}(t) = (1-\alpha)\tilde{D}(t-1) + \alpha D(t) \quad (1)$$

$$\tilde{D}(t) > D_{min} + (D_{max} - D_{min})\delta \quad (2)$$

where $\delta \in (0,1)$ is a custom threshold parameter. Throughout this paper, we use the values $\alpha = 1/8$, $\delta = 0.15$ that are selected in [5] by means of simulation experiments.

In absence of early-congestion indication, LP behaves like standard TCP Reno, i.e., performing an additive increase of the congestion window $cwnd$ which can easily be gathered from Fig. 1-(b) and (f). Whenever an early congestion is detected, according to the rules outlined above, LP halves the congestion window and enters an inference phase by starting an inference timeout timer. During this period, LP only observes responses from the network and avoids increasing the congestion window. After this phase, if congestion persists it reduces the congestion window to zero and restarts the TCP

Reno congestion avoidance scheme. Finally, in case of losses, LP behaves like TCP Reno.

*B. TCP-NICE*

TCP-NICE (or NICE *tout court*) instead maintains a minimum $RTT_{min}$ and maximum $RTT_{max}$ estimates of the round trip delay. Congestion is detected when more than a given fraction $\phi$ of packets during the same RTT experiences a delay exceeding:

$$RTT > RTT_{min} + (RTT_{max} - RTT_{min})\delta \quad (3)$$

where $\delta$ and $\phi$ are protocol parameters set to $\delta = 0.2$ and $\phi = 0.5$ as in [6]. Notice that (3) is the same formula of LP (1), but computed on the RTT variable, and using the fraction-trick instead of a moving average.

In the absence of congestion, NICE behaves like TCP-Vegas [20], whose congestion window dynamics are delay-based (and thus rather different from loss-based dynamics). Whenever early-congestion is signaled, NICE simply halves its congestion windows and sending rate, practically reintroducing the multiplicative decrease behavior. Finally, when a loss is detected it instead reacts like TCP Reno.

The fact that NICE inherits its congestion control behavior from Vegas [20] rather than from TCP Reno has profound impact on the *cwnd* evolution: as can be gathered from Fig. 1-(c) and (g), NICE shows a much smoother behavior as its throughput stabilizes around the effective link capacity. We point out that NICE allows *cwnd* to be a fraction of 1 by sending one packet after waiting for the appropriate number of RTTs: the use of fractional values for *cwnd* guarantees non-intrusiveness even in the case of many NICE flows sharing the same bottleneck.

*C. LEDBAT*

Finally, LEDBAT maintains a minimum one-way delay estimation $D_{min}$, which is used as base delay to infer the amount of delay due to queuing. LEDBAT flows have a target queuing delay $\tau$, i.e., they aim at introducing a small, fixed, amount of delay in the queue of the bottleneck buffer. Flows monitor variations of the queuing delay $D(t) - D_{min}$ to evaluate the distance $\Delta(t)$ from the target as in (4):

$$\Delta(t) = \tau - (D(t) - D_{min}) \quad (4)$$
$$cwnd(t+1) = cwnd(t) + \gamma\Delta(t)/cwnd(t) \quad (5)$$

where $\tau, \gamma$ are protocols parameters that we study later on.

In absence of early-congestion indication, i.e., when the target $\tau$ has not been reached yet, $\Delta(t) > 0$ in (4) and thus *cwnd* grows as defined by (5). Notice that when the target is reached, $\Delta(t) = 0$ thus *cwnd* settles.

Values of $\Delta(t) < 0$ are perceived as early-congestion indication (i.e., other traffic is increasing the queuing delay $D(t) - D_{min}$), to which LEDBAT reacts by reducing *cwnd* proportionally to the offset from the target according to (5). Finally, in case of losses, it behaves like TCP Reno.

Overall, LEDBAT shares similarities with, and exhibit differences from, the other LBE protocols: (i) as LP, it does rely on one-way delay estimates to detect congestion, but unlike LP it does not employ a smoothing average; (ii) as NICE, its congestion controller is based on the delay, but unlike NICE it employs a PID controller in order to reach (or deviate from) the target delay. As can easily be gathered from Fig. 1, the behavior of LEDBAT is however closer to NICE than to LP.

III. METHODOLOGY

*A. Simulation scenarios*

We employ `ns2` simulations to address the comparison of LBE protocols. While TCP Reno and LP protocols are already implemented, we implement both NICE and LEDBAT congestion control protocols in the network simulator. Source code of our LEDBAT implementation can be found at [12]. As reference network scenario, we use a dumbell topology where the capacity of the bottleneck is fixed to $C = 10$ Mbps, the one-way propagation delay equals 25 ms (thus round trip delay is equal to $RTT = 50$ ms), and the buffer size is set to $B_{max} = 100$ packets. We consider backlogged sources[1], that use a fixed packet size equal to $S =$ 1500 Bytes. All TCP and LBE sources start simultaneously, so that we avoid potential late-comer issues [13], and last for 120 seconds.

In this work we first focus on a sensitivity analysis of LEDBAT, to assess the impact of parameters $\tau$ and $\gamma$ on the system performance. We carry on this analysis in both (i) an inter-protocol case, where a TCP Reno flow and a LEDBAT flow share the bottleneck and (ii) an intra-protocol case, where two LEDBAT flows compete against each other. The aim of (i) is to determine whether $\tau$ and $\gamma$ offer the chance to tune the level of LBE priority in LEDBAT, while (ii) aims at verifying whether unfairness may arise among legacy LEDBAT implementations (e.g., different releases of the same code, different implementations or parameter settings, etc.).

We then focus on a comparison of TCP and LBE protocols, again considering two cases: (iii) a single TCP flow shares the bottleneck with a varying number of homogeneous LBE flows (i.e., same LBE protocol) and (iv) several heterogeneous LBE flows compete against each other. In both cases, our aim is to evaluate the level of low priority of LBE protocols. Finally, we consider more realistic scenarios in (v) by taking into account the impact of RTT heterogeneity on LBE performance.

*B. Evaluation metrics*

Performance evaluation is carried out considering different metrics, that relate to either network-centric (e.g., efficiency, average queue size) or user-centric performance (e.g., fairness, packet loss rate). For the sake of example, Fig. 1-(a) summarizes the performance of flows in corresponding scenarios in terms of some of these metrics (i.e. efficiency, fairness, and breakdown).

*Bottleneck link efficiency ($\eta$)* is the primary network-centric metric, and expresses the link utilization as the ratio between the sum of the throughput values $x_i$ achieved by all flows over the available capacity $\eta = \sum_i x_i / C$.

---
[1] As we consider backlogged sources only, dynamics of LEDBAT are well described by means of (5) only; in case of non-backlogged sources, the dynamics changes slightly to avoid cwnd increase indefinitely [1]

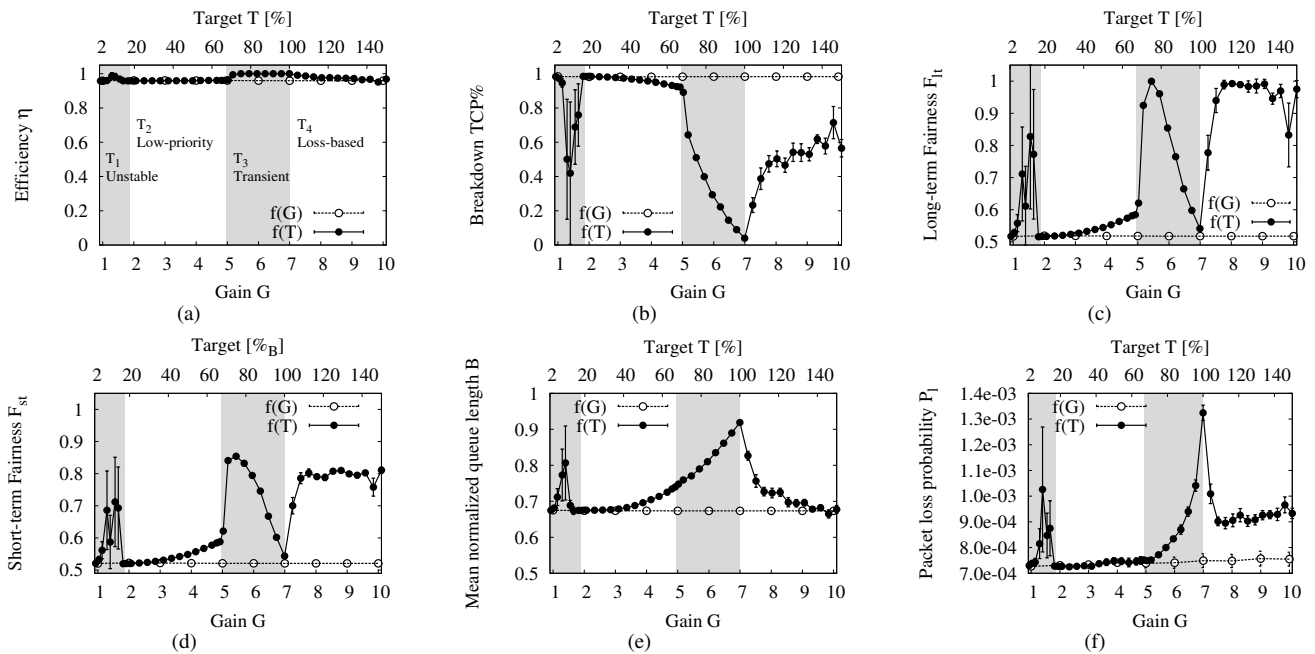

Fig. 2. LEDBAT vs TCP Reno: Inter-protocol sensitivity analysis, for varying LEDBAT target $T$ and gain $G$ parameters

*Average queue occupancy index (B)* is computed averaging buffer occupancy during the simulation (measured at each enqueue event in the buffer), and normalizing the value over the buffer size $B = E[B]/B_{max}$ for convenience.

Whenever the buffer overruns and packets are dropped, all protocols drastically reduce their sending window: *packet loss probability ($P_l$)* therefore relates to user-performance, and is computed as the ratio of the dropped packets over the total number of packets sent on the link.

We further express the system performance using two metrics apt at describing how the link resources are shared among flows. To gauge the impact of LBE on TCP, we define *TCP breakdown ($TCP_\%$)* as the TCP Reno traffic share percentage over the total amount of data exchanged on the link, i.e., $TCP_\% = \sum_{j \in TCP} x_j / \sum_i x_i$.

We further describe the capacity share in terms of *Jain fairness index (F)*, defined as $F = (\sum_{i=1}^{N} x_i)^2 / (N \cdot \sum_{i=1}^{N} x_i^2)$ where $N$ is the number of considered flows and $x_i$ is the rate of the $i-th$ flow. In the best case, when all flows get a fair share of the resources, $F$ is equal to one, while in the worst one, namely when a single flow exploits all the link, it is equal to $1/N$.

We compute the fairness index over both the whole flow duration and over a smaller time scales (considering a temporal window of 20 RTT, or equivalently 1 s): we refer to *long-term fairness ($F_{lt}$)* in the first case, and to *short-term fairness ($F_{st}$)* in the latter one. Notice that the ability to achieve short-term (vs long-term) fairness may have rather different implications, e.g., if we consider the case of several P2P flows measuring throughput to perform peer selection (as long-term fairness may not be sufficient and significantly bias peers decisions).

## IV. LEBDAT SENSITIVITY ANALYSIS

### A. Inter-protocol: LEDBAT vs TCP

We start our sensitivity analysis by considering two flows, a standard TCP Reno flow and a LEDBAT one, that start simultaneously and vary, one at the time, the values of parameters $\tau$ and $\gamma$. Notice that the standardization draft *does not specify any value* for the gain parameter $\gamma$. Conversely, the draft *specifies a mandatory value* for the target parameter equal to $\tau = 25$ ms. This choice of $\tau$ is somewhat arbitrary, and based on experimental observations (whose results are however unreported so far) or motivated by practical constraints (e.g., today limits in the precision of the delay measurement, etc.), rather than being based on concrete foundations. As such, $\tau$ is often referred to as "magic number" with a deprecatory sense in LEDBAT WG discussion [21]: therefore, we believe that a thorough exploration of the impact of the above parameters is necessary, which we carry on by simulation.

For convenience, we re-express the gain parameter $\gamma$ as multiples of the target $\tau$, i.e. $G = \gamma \tau$, and explore the range $G \in [1, 10]$. We also re-express the target delay parameter $\tau$ in terms of buffer percentage as $T = \tau C / (SB_{max})$, and explore the range $T \in [2, 150]\%$, corresponding to $\tau \in [2.4, 180]$ ms. For reference purposes, notice that the mandatory draft value $\tau = 25$ ms correspond to $T = 20\%$, while a full buffer occupancy $T = 100\%$ is attained when $\tau = 120$ ms.

Fig. 2 reports the simulation results for each of the metric $f \in \{\eta, TCP_\%, F_{st}, F_{lt}, B, P_l\}$ described early in Sec. III-B, arranged as one per plot. In each plot, we report two curves, namely $f(G)$ and $f(T)$. The $f(G)$ curve reports how $f(\cdot)$ varies as a function of the gain $G \in [1, 10]$ (on the bottom x-axis), when target is fixed to $\tau = 25$ ms. The $f(T)$ curve instead reports how $f(\cdot)$ varies as a function of the target $T \in [2, 150]\%$ (on the top x-axis), when gain is fixed to $G = 1$.

From all the subplots we can see that, for all metrics, the $f(G)$ curve is roughly flat, i.e., the gain parameter only minimally affects the behaviour of the LEDBAT protocol in this case. This can be explained with the fact that, as LEDBAT is designed to yield to TCP, it will yield irrespectively of $G$. The gain value thus only affects the speed at which LEDBAT will yield, which thus quickly happens for any value of $G$.

Therefore, from now on we restrict our attention the impact of the target parameter, and analyze the behavior of the $f(T)$ curves. In Fig. 2-(a) we can see that the efficiency $\eta$ is only slightly influenced by the variation of the target and remains always close to the total link capacity. This is expected, as even if the target is misconfigured, either LEDBAT or TCP Reno can take advantage of the unused bandwidth, which result in an overall efficient use of the link capacity.

Considering instead the $TCP_\%$ reported in Fig. 2-(b), we can identify four working regions. When the target is very small $T_1 \in [2, 18]\%$ the LEDBAT protocol is not always able to reach the target delay, which leads to shaky $TCP_\%$ behavior. In a second region $T_2 \in [18, 65]\%$, LEDBAT completely yields to the TCP Reno flows, working in low-priority mode and thus attaining its goal. In a third region $T_3 \in [65, 100]\%$, LEDBAT aggressively start to erode bandwidth to the TCP Reno flow: this causes losses in the TCP Reno flow, which progressively backoff; as a consequence, the $TCP_\%$ starts decreasing until LEDBAT has the monopoly of the buffer when $T = 100\%$ and TCP Reno starves ($TCP_\% \simeq 0\%$). Finally, in the fourth region $T_4 > 100\%$ the target exceeds the buffer size: in this case, LEDBAT falls in the TCP Reno-like loss-based behavior, increasing the sending rate until a loss occurs, which immediately drop down its rate. As a consequence, the breakdown is now more similar ($TCP_\% \simeq 50\%$)

Similar considerations can be gathered by looking at the long-term $F_{lt}$ or short-term $F_{st}$ fairness plots shown in Fig. 2-(c) and Fig. 2-(d) respectively: indeed, to an even breakdown correspond maximum fairness ($F_{st} \simeq 1$) while to an uneven breakdown, favoring either TCP Reno ($TCP_\% \simeq 100\%$) or LEDBAT ($TCP_\% \simeq 0\%$), always correspond minimum fairness values ($F_{st} \simeq 1/2$). We also notice that, although as expected short-term fairness is more difficult to achieve ($F_{st} < F_{lt}$), the same qualitative behavior holds for both fairness timescales.

From Fig. 2-(e) and Fig. 2-(f) we see that, as expected, increasing target the average buffer occupancy increases besides the occupancy due to TCP Reno. Losses increase as well reaching a peak for $T = 100\%$, corresponding to LEDBAT maximum aggressiveness; afterward LEDBAT is in loss-mode, and the scenario degenerates into two TCP Reno flows sharing a bottleneck.

Overall, the sensitivity analysis suggests that, although LEDBAT spans a wide range of low-priority levels (especially in the third region), its tuning is highly impractical. Indeed, the support of target values $T_3 \in [65, 100]\%$ is very small, meaning that small variation of $T$ lead to completely different scenarios, where either LEDBAT or TCP Reno exhibit starvation. Moreover, the actual values of $\tau$ yielding to a specific level of low-priority depends on network parameter (e.g., capacity $C$, buffer size $B$) and are likely affected from other factors as well (e.g., number of TCP Reno flows, heterogeneous RTT, etc.)

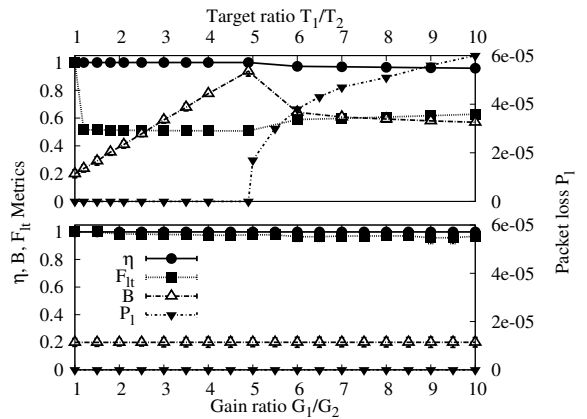

Fig. 3. LEDBAT vs LEDBAT: Intra-protocol sensitivity analysis, for varying LEDBAT target $T_1/T_2$ and gain $G_1/G_2$ ratios

*B. Intra-protocol: LEDBAT vs LEDBAT*

We pursue our sensitivity analysis by considering two LEDBAT flows with heterogeneous settings sharing the same bottleneck link. We perform two sets of experiments, varying either (i) the gain ratio $G_1/G_2$ of the two flows when $G_2 = 1, \tau = 25$, or (ii) the target ratio $T_1/T_2$ when $T_2 = 20\%, \gamma = 1/\tau$. In both cases, the ratio varies in the $[1, 10]$ range. Results of the sensitivity analysis are reported in Fig. 3, that depicts the packet loss rate $P_l$ (right y-axis), the average buffer size $B$, the efficiency $\eta$ and the fairness $F_{lt}$ (left y-axis) as a function of the $T_1/T_2$ target ratio (top plot) and $G_1/G_2$ gain ratio (bottom plot).

As in the previous case, it is easy to gather that impact of gain is very modest, even in the case of a 10-fold factor. This phenomenon has an intuitive explanation. Consider indeed, that flow with the largest gain will start moving faster that the other flow toward the target. However, after the first flow increases its window, the convergence speed toward target will slow down, since the differences between the target and the measured delay is now smaller for the first flow than for the second. In other words, the difference in the delay offset in (5) compensates for differences in the gain factor $\gamma$.

Conversely, even slight differences in the target settings may have strong consequences as can be seen in the top of Fig. 3. Indeed, as soon as $T_1/T_2 > 1$ it can be seen that the fairness immediately drops to its minimum value $F_{lt} = 0.5$. This is due to the fact that flows with higher-target are always more greedy than their lower-target counterpart. As a matter of fact, if both flows start at the same time, they both measure the same base delay, and the higher-target flow converges faster to its target and stabilizes: as the amount of queuing is now larger than that of the less aggressive flow, this one back offs and starves. This hold until $T_1 + T_2 > 100\%$ (which happens when $T_1/T_2 = 5$ given that $T_2 = 20\%$), in which case both LEDBAT flows may experience packet drops: nevertheless, higher-target flow will always be advantaged prior that losses occur, and so the unfairness persists.

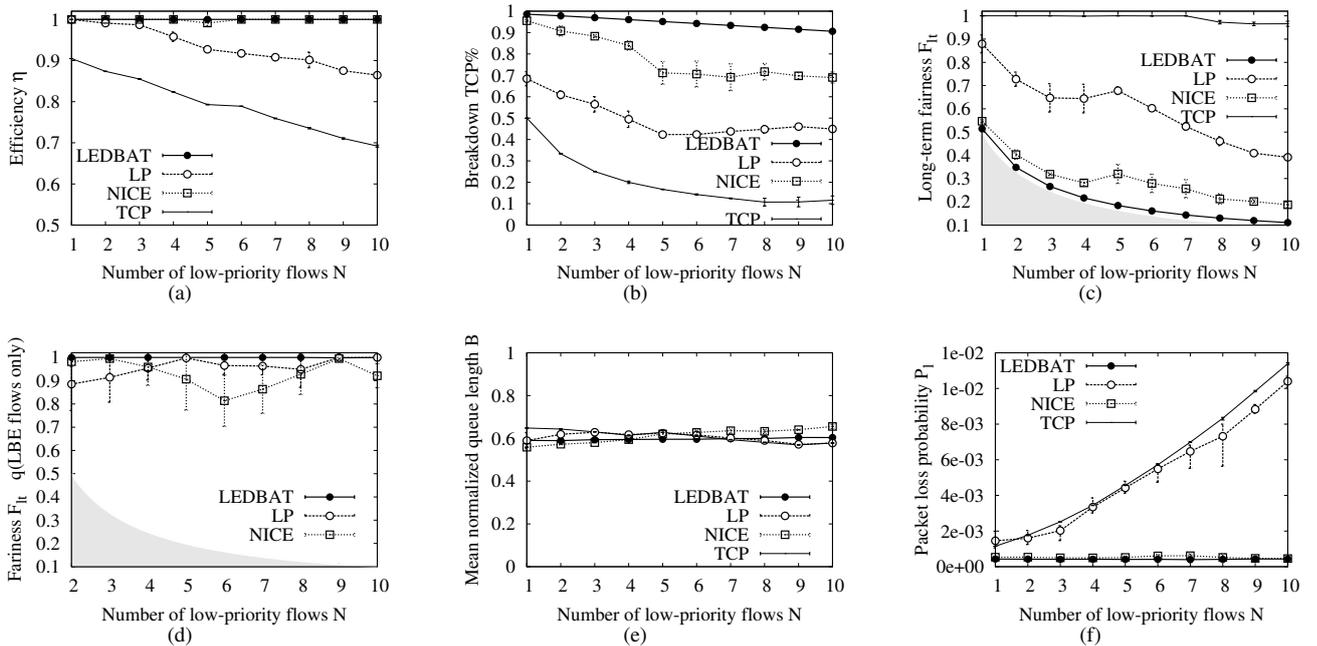

Fig. 4. LBE against one TCP flow: Impact of the number LBE flows on the system performance

Overall, we see that gain and target parameters have rather different effects: on the one hand, provided that LEDBAT flows have the same target, differences in gain do not entail any unfairness among flows. On the other hand, even small difference in targets yield to extremely unfair situation: this is a delicate point, which we believe deserves further attention.

## V. LBE PROTOCOLS COMPARISON

### A. LBE against TCP

We now fix LEDBAT parameters and consider a larger set of LBE protocols in the comparison. Following our sensitivity analysis, we know that $\gamma$ selection is less critical than $\tau$ one: we set $\gamma = 1/\tau$, and use the mandatory value $\tau = 25\,\text{ms}$ which we verified to be a robust choice.

We consider a typical scenario where $N$, ($N \in [1, 10]$) low-priority flows (e.g., due to P2P or other service) share the same bottleneck with a single TCP Reno connection, (representative of a generic high-priority service), for a total of $N + 1$ flows. We perform several set of simulations separately, considering each time a different LBE protocol. For reference purpose, we also simulate the case where $N + 1$ TCP Reno flows share the same bottleneck.

Results for the common set of metrics are reported in Fig. 4. Considering efficiency $\eta$ in Fig. 4-(a), we see that delay-based NICE and LEDBAT are able to fully utilize the spare bandwidth left by TCP Reno. Conversely, in the LP or TCP Reno cases, losses entail an efficiency reduction (especially for large $N$).

Breakdown $TCP_\%$ reported in Fig. 4-(b), states that e.g., in the $N = 10$ LEDBAT case, TCP Reno consumes about 90% of the link capacity (since $\eta \simeq 1$), leaving thus the $N = 10$ LEDBAT flows a mere 1% of the capacity each. Comparing this result with NICE (about 3% each) or LP (about 5% each) under the same $N = 10$ settings, we gather that LEDBAT achieves the lowest priority, closely followed by NICE. This is further exacerbated from the long-term fairness plot of Fig. 4-(c), showing that in the LEDBAT and NICE cases fairness approaches the minimum possible value (i.e., the shaded region indicates values that fairness cannot achieve since they are smaller than the lower bound $\frac{1}{N+1}$ for the fairness index).

The plot in Fig. 4-(d) depicts instead the long-term fairness $F_{lt}$ evaluated over the $N$ LBE flows only. It can be seen that fairness is always high, meaning that generally the excess that remains after the TCP breakdown of Fig. 4-(b), is evenly shared among LBE flows. Notice that fairness among LBE flows is however lower in the case of NICE, where apparently some LBE flow opportunistically take advantage of the others.

Finally, average occupancy index $B$ and packet loss $P_l$ are reported in Fig. 4-(e) and (f) respectively. Again, delay-based versus loss-based congestion control principles are remarkably different, which is especially true in case of the loss curve: interestingly, despite its low priority aim, the amount of loss induced by LP is strikingly similar to that of classic TCP Reno. Delay-based versus loss-based difference, although less evident, also reflects on the queue size: indeed, TCP Reno and LP average queue size decrease when losses increase along with the number of flows; conversely, queue occupancy in the NICE case slowly arises for increasing $N$, and is practically unaffected by $N$ in the LEDBAT case.

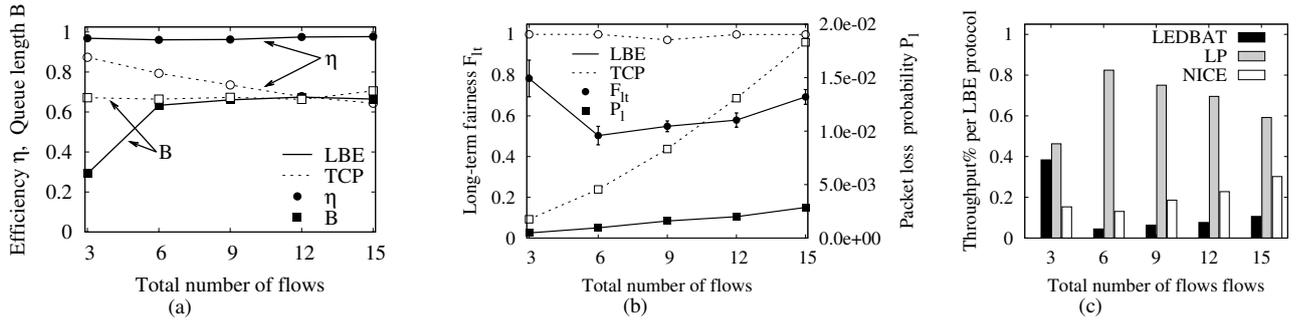

Fig. 5. LBE against LBE: LP, LEDBAT and NICE competing for the same bottleneck, compared with the same number of TCP Reno only flows.

### B. LBE against LBE

In order to investigate the mutual interaction of the different lower-priority protocols, we define an heterogeneous scenario in which several LEDBAT, LP and NICE flows contend the same bottleneck link. We perform different tests in which an increasing number of flows is considered, from 1 to 5 for each flavor (which corresponds to a total of 3 to 15 flows). As reference, we perform also the corresponding experiment with the same number of TCP Reno flows only (i.e., 3 to 15 TCP Reno flows). We choose for all the LEDBAT flows, the standard parameters values, namely $\tau = 25\,ms$ and $\gamma = 1$. We point out that qualitatively similar results can be gathered using different parameter settings, which we are however unable to report for lack of space.

Let us start by examining the efficiency $\eta$ and average normalized buffer length $B$, which are reported in Fig. 5-(a). Looking at the efficiency metric, we can see that in the heterogeneous LBE scenario, flows are able to utilize the available resource fully, with $\eta$ always close to its maximum. On the contrary, the efficiency in the case of all TCP Reno flows progressively decrease as long the number of competing flow increase, due to the typical synchronization behavior of the protocol after loss. Looking at the normalized average queue size we can notice that the average $B \simeq 2/3$ is not affected by the number of flows in the TCP Reno case. In the LBE case instead, average queue size approaches that of TCP Reno only when at least two flows per protocol insist on the bottleneck. When only a total of three LBE flows are competing for the resource, a rather unexpected phenomenon arises: in this case, LEDBAT often forces LP in low-priority mode and is thus able to exploit a significant part of the resource. As a consequence, the average queue size $B$ reflects the LEDBAT target $\tau$, plus the contributions due to LP and NICE. When more than two LP flows are instead present on the bottleneck, their behavior synchronize and is perceived as more aggressive by LEDBAT: in this case, it is more rare that both LP flows goes into inference mode at exactly the same time, thus LEDBAT has fewer opportunities to profit of the resource.

Packet loss probability $P_l$ and long-term fairness $F_{lt}$ are instead reported in Fig. 5-(b). Concerning packet loss, since 2/3 of the total flow number consists of delay-based protocols, the loss rate is clearly lower than the TCP Reno reference case. Long-term fairness performance shown in Fig. 5-(b) is instead better understood by considering also the throughput breakdown reported in Fig. 5-(c), in which each bar represents the percentage of traffic due to a particular LBE protocol. As expected, fairness between heterogeneous LBE flows is lower than that of homogeneous TCP Reno connections, but is however higher than the LBE-TCP Reno performance early reported in Fig. 4-(c). In particular, maximum LBE fairness is achieved when only one flow per each LBE flavor is considered: from Fig. 5-(c) we see that LP and LEDBAT performance are very close in this case, which raises the fairness metric. When the number of considered low-priority flow increases, the fairness instead decreases due to a higher aggressiveness of the LP protocol.

### C. Impact of the RTT heterogeneity

Finally, we report on the impact of RTT heterogeneity in Fig. 6. We consider two flows, of the same protocol type (LBE or TCP) sharing the same bottleneck, that have a different round trip delay expressed by the $RTT_1/RTT_2$ ratio. We perform simulations separately for each protocol, exploring the $RTT_1/RTT_2 \in [1, 10]$ range; $RTT_1$ is increased by adding propagation delay to the return path, so that one-way delay estimation on the forward path are not affected. Top plot of Fig. 6 reports the long term fairness $F_{lt}$, while bottom plot reports the efficiency $\eta$ as a function of the RTT ratio.

Interesting remarks can be gathered from the plots. Concerning the fairness metric, it can be seen that only NICE, by virtue of its inheritance of Vegas [20] congestion control, provides fairness in case of heterogeneous RTT settings. However, this comes at the price of a reduced efficiency, since in order to be fair, the more aggressive small-RTT flow has to slow down its rate to match that of the large-RTT flow. Efficiency loss happens also in the case of LP and TCP Reno, despite they are unable to offer fairness either. Finally, LEDBAT realized a perfectly efficient system, which comes at the price of an unfair share of the resources: although the congestion control works only on the forward path, due to the fastest feedback, the small-RTT flow is able to reach its target first, whereas the second flow will see that a queuing delay (due to the small-RTT flow) equals to its target, and will thus settle in a starvation state.

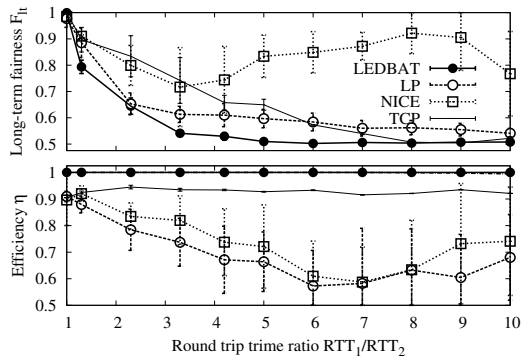

Fig. 6. Impact of the RTT heterogeneity

## VI. CONCLUSION

This paper analyzed different Lower-than Best Effort (LBE) transport protocols behavior: by means of simulation, we carried on a thorough comparison of LEDBAT, LP and NICE, studying the impact they have on TCP Reno traffic, as well as their mutual impact. From our sensitivity analysis of LEDBAT we gather that it is hard to tune its behavior, and especially its level of priority with respect to TCP Reno by means of a simple adjustment of its gain $\gamma$ and target $\tau$ parameters. Indeed, the gain has practically no influence, while the impact of target can not be controlled, as changes in the system performance are too steep. Also, we see that gain and target parameters have rather different effects if we consider the coexistence of legacy LEDBAT flows with heterogeneous settings: on one hand, provided that LEDBAT flows have the same target, differences in *gain* do not entail any unfairness among flows; on the other hand, even small *target* difference yield to extremely unfair situation. From this part of the analysis we conclude that tuning LEDBAT is thus a delicate point, which deserves further attention in the future, which holds true even when heterogeneous network settings (e.g., RTT) are considered.

From our comparison study, we gather that LEDBAT achieves the lowest possible priority with respect to NICE and LP. Moreover, we find that LP inherits from its loss-based design a higher aggressiveness than the delay-based NICE whose degree of low-priority sits thus in between LEDBAT and LP. Interestingly, we point out that there are also limit cases (e.g., only an LP, LEDBAT and NICE flows sharing the same bottleneck) in which the low-priority degree can exhibit unexpected behavior (i.e., as LEDBAT is in this case as aggressive as LP).

We believe this work makes a first important step in understanding, comparing and ranking several LBE protocols. At the same time, an important question remains open: namely, how a different degree of low-priority can be achieved in a robust, tunable fashion, which our future research aims at answering.

ACKNOWLEDGEMENT

This work was supported by Celtic Project TRANS.

REFERENCES

[1] S. Shalunov. (2010) Low Extra Delay Background Transport (LEDBAT). [Online]. Available: http://tools.ietf.org/id/draft-ietf-ledbat-congestion-01.txt
[2] A. Norberg. uTorrent transport protocol. BitTorrent Enhancement Proposals. [Online]. Available: http://www.bittorrent.org/beps/bep_0029.html
[3] "Comcast throttles bittorrent traffic, seeding impossible," August 2007. [Online]. Available: http://torrentfreak.com/comcast-throttles-bittorrent-traffic-seeding-impossible/
[4] (2007) Setting up bittorrent with qos. [Online]. Available: http://forums.whirlpool.net.au/forum-replies-archive.cfm/627285.html
[5] A. Kuzmanovic and E. Knightly, "TCP-LP: low-priority service via end-point congestion control," *IEEE/ACM Transactions on Networking (TON)*, vol. 14, no. 4, p. 752, 2006.
[6] A. Venkataramani, R. Kokku, and M. Dahlin, "TCP Nice: A mechanism for background transfers," *ACM SIGOPS Operating Systems Review*, vol. 36, no. SI, p. 343, 2002.
[7] S. Liu, M. Vojnovic, and D. Gunawardena, "4CP: Competitive and considerate congestion control protocol," 2006.
[8] Microsoft background intelligent transfer service (bits). [Online]. Available: http://msdn.microsoft.com/en-us/library/aa363167.aspx
[9] E. Altman, K. Avrachenkov, and B. Prabhu, "Fairness in MIMD congestion control algorithms," *Telecommunication Systems*, vol. 30, no. 4, pp. 387–415, 2005.
[10] Y. Li, D. Leith, and R. Shorten, "Experimental evaluation of TCP protocols for high-speed networks," *IEEE/ACM Transactions on Networking (ToN)*, vol. 15, no. 5, p. 1122, 2007.
[11] S. Molnár, B. Sonkoly, and T. Trinh, "A comprehensive TCP fairness analysis in high speed networks," *Computer Communications*, vol. 32, no. 13-14, pp. 1460–1484, 2009.
[12] Ledbat implementation for ns2. [Online]. Available: http://perso.telecom-paristech.fr/~valenti/pmwiki/pmwiki.php?n=Main.LEDBAT
[13] D. Rossi, C. Testa, S. Valenti, P. Veglia, and L. Muscariello, "News from the internet congestion control world," Technical Report, Aug 2009.
[14] D. Rossi, C. Testa, and S. Valenti, "Yes, we LEDBAT: Playing with the new BitTorrent congestion control algorithm," in *In PAM'10*, Zurich, Switzerland, Apr 2010.
[15] D. Qiu and R. Srikant, "Modeling and performance analysis of BitTorrent-like peer-to-peer networks," in *ACM SIGCOMM'04*, Portland, Oregon, USA, Aug 2004.
[16] A. R. Bharambe, C. Herley, and V. N. Padmanabhan, "Analyzing and Improving a BitTorrent Performance Mechanisms," in *IEEE INFOCOM'06*, Barcelona, Spain, Apr 2006.
[17] R. Bindal, P. Cao, W. Chan, J. Medved, G. Suwala, T. Bates, and A. Zhang, "Improving Traffic Locality in BitTorrent via Biased Neighbor Selection," in *IEEE ICDCS '06*, Lisboa, Portugal, Jul 2006.
[18] M. Izal, G. Urvoy-Keller, E. W. Biersack, P. Felber, A. Al Hamra, and L. Garcés-Erice, "Dissecting BitTorrent: Five Months in a Torrent's Lifetime," in *In PAM'04*, Antibes, France, Apr 2004.
[19] V. Jacobson, "Congestion avoidance and control," *ACM SIGCOMM*, vol. 25, no. 1, 1988.
[20] L. Brakmo, S. O'Malley, and L. Peterson, "TCP Vegas: New techniques for congestion detection and avoidance," *ACM SIGCOMM Computer Communication Review*, vol. 24, no. 4, pp. 24–35, 1994.
[21] LEDBAT Mailing List Archives. [Online]. Available: http://www.ietf.org/mail-archive/web/ledbat